\documentclass[aps,amsfonts,prl,twocolumn,showpacs]{revtex4}
\usepackage{amsmath,amssymb,bm,epsfig,graphics,verbatim}
\usepackage[usenames,dvipsnames]{color}
\def\nop{{\mathbf Q}}
\def\ncp{{\mathbf Y}}

\def\LDLa{\mathcal{A}}
\def\smear{{\mathrm H}}
\def\ipbra{\{}
\def\ipket{\}}

\def\stiff{\mathcal{L}}

\def\mycorrel{{\cal C}}

\def\correllength{\xi}
\def\loclen{\correllength_{L}}

\newcommand{\be}{\begin{equation}}
\newcommand{\ee}{\end{equation}}
\newcommand{\ba}{\begin{eqnarray}}
\newcommand{\ea}{\end{eqnarray}}
\newcommand{\bw}{\begin{widetext}}
\newcommand{\ew}{\end{widetext}}

\newcommand{\rv}{{\bm{r}}}

\begin{document}

\title{Phenomenological Theory of Isotropic-Genesis Nematic Elastomers}

\author{Bing-Sui~Lu$^1$}
\author{Fangfu Ye$^2$}
\author{Xiangjun Xing$^3$}
\author{Paul M.~Goldbart$^2$}
\affiliation{$^1$Department of Physics,
University of Illinois at Urbana-Champaign, 1110~West Green Street, Urbana, IL~61801}
\affiliation{$^2$School of Physics, Georgia Institute of Technology,
837~State Street, Atlanta, GA~30332}
\affiliation{$^3$Department of Physics and Institute of Natural Sciences, Shanghai Jiao Tong University, Shanghai, China}

\date{\today}

\begin{abstract}
We consider the impact of the elastomer network on the structure and fluctuations in the isotropic-genesis nematic elastomer, via a phenomenological model that underscores the role of network compliance.  The model contains a network-mediated nonlocal interaction as well as a new kind of random field, which reflects the memory of the nematic order present at cross-linking, and also encodes local anisotropy due to localized polymers.  Thus, we predict a regime of short-ranged oscillatory spatial correlations (both thermal and glassy) in the nematic alignment trapped into the network.
\end{abstract}

\maketitle
Consider a melt or solution of nematogenic polymers, by which we mean long, flexible polymers carrying rod-like units. These units, which give the system the possibility of exhibiting liquid crystallinity, may be integrated along the polymer chain backbones (the main-chain case) or in groups that dangle from the backbone (the side-chain or pendant case).  Now consider the process of instantaneous cross-linking.  Here, one begins with the melt or solution at equilibrium and---so rapidly that hardly any relaxation has time to occur---one introduces permanent bonds between some random fraction of the pairs of chain segments that happen, at the instant of cross-linking, to be nearby one another.  When the cross-linking process is carried out in the {\it isotropic\/} state of the nematogens, the resulting material is called an {\it isotropic-genesis nematic elastomer\/} (or IGNE; see, e.g.,~Ref.~\cite{LCE:WT,urayama,urayamaPMtransition,terentjev1,terentjev2,uchida,selinger}).
It is a macroscopic random network medium that \lq\lq memorizes\rq\rq\ both the positions of the chain segments and the orientations of the nematogen units at the instant of cross-linking.  The memorization is, however, only partial, as a result of the thermal fluctuations that occur in the new, post-cross-linking equilibrium state.

To describe the structure and correlations of the system post cross-linking, we employ the local nematic order parameter $Q_{dd^{\prime}}({\bf r})$, which is traceless, symmetric, and of rank-two, and is defined microscopically via
\begin{equation}
Q_{dd^{\prime}}({\bf r})=\sum_{p=1}^{P}
\big(N_{d}^{p}N_{d^{\prime}}^{p}-D^{-1}\delta_{dd^{\prime}}\big)\,
\delta^{(D)}\big({\bf r}-{\bf R}^{p}\big),
\end{equation}
where $P$ is the number of rod-like units,
${\bf N}^{p}$ is the microscopic unit orientation vector of unit $p$ and
${\bf R}^{p}$ is its microscopic position vector in $D$-dimensional space, and
${\bf r}$ is an arbitrary position vector.
In addition, we characterize the random local environmental anisotropy, which tends to induce local nematic alignment $\nop$ in the post cross-linking system, via the tensor $\ncp_t$:
\begin{equation}
\label{eq:anisotropy}
\ncp_t({\bf r}) = \ncp({{\bf r}})+
\frac{T}{T_p} \int d^{D}r^{\prime}\,
{\smear}({\bf r}-{\bf r}^{\prime})\,
\nop^{0}({\bf r}^{\prime}).
\end{equation}
This random environmental anisotropy is caused by the thermally averaged part of random local spatial arrangement of the localized polymers at post-cross-linking equilibrium.  It consists of two parts: (i)~a part that is independent of the the pattern of local nematic alignment $\nop^{0}$ present at the instant of cross-linking, which we call the {\it memory-independent random field\/} and denote by $\ncp$; and (ii)~another part that is due to the pattern of $\nop^{0}$, which we call the {\it memory-dependent random field\/}.  The pattern $\nop^{0}$ is partially imprinted in the network structure, and this imprint then partially elicits a response similar to $\nop^{0}$ in the post-cross-linking state.  The relationship between $\nop$ and $\nop^{0}$ is characterized by a \lq\lq smearing\rq\rq\ kernel, which embodies the idea that $\nop$ (i.e., the post-cross-linking equilibrium-state memory of $\nop^{0}$) is partially erased, as a result of the positional thermal fluctuations of the network.  Equivalently, viewed from wave-vector space, the contribution from $\nop^{0}$ becomes ${\smear}^{\phantom{0}}_{\bf k}\nop^{0}_{\bf k}$.  Physically, we expect $\smear(\rv) $ to be positive and bell-shaped, operative primarily over a region of order the typical localization length $\loclen$ (which reflects how weakly localized the network constituents are; see, e.g., \cite{Vulcan_Goldbart}), and to decay monotonically with increasing $\vert\rv\vert$ over this lengthscale, ultimately tending to zero for $\vert\rv\vert\gg\loclen$.  Correspondingly, in wave-vector space ${\smear}_{\bf k}$ would decay monotonically to zero over a scale $\loclen^{-1}$.  Hence, we see that ${\smear}$ serves as a
\lq\lq soft filter\rlap,\rq\rq\
de-amplifying---more strongly the shorter the lengthscale---the contributions made by the Fourier components of $\nop^0$ to the random anisotropic environment on distance scales shorter than $\loclen$.  This is a natural consequence of the liquid-like character of the post-cross-linking system on lengthscales shorter than $\loclen$.
As for the overall {\it amplitude\/} of $\smear$, this we expect to {\it increase\/} with
(i)~the fraction $G$ of polymers that are localized,
(ii)~the sharpness of localization, $1/\xi_L$,
(iii)~the nematogen-nematogen aligning interaction $J$, and
(iv)~the length $\ell$ of the nematogens; and
we expect it to {\it decrease\/} with the \lq\lq measurement temperature\rq\rq\ $T$
(see below for more on this concept),
because thermal fluctuations tend to moderate any aligning forces.
A complementary microscopic calculation~\cite{ref:microscopic} bears out this expectation, yielding
$\smear_{\bf k}=\smear_{\bf 0}\exp(-k^2\xi_L^2/2)$,
where the amplitude $\smear_{\bf 0} = G^{2}J^{2}(\ell/\loclen)^{4}/T$.

In terms of these ingredients, we take as a model for the Landau-type free-energy cost $F$ associated with the induction of local nematic order in the post-cross-linking system the form:
\begin{eqnarray}
\label{eq:landau}
F&=&
\frac{1}{2}\int_{{\bf k}}
\Big(
\big(
\LDLa t+\stiff k^{2}+{\smear}_{\bf k}
\big)
\big\ipbra \nop_{{\bf k}}\nop_{-{\bf k}}\big\ipket
\nonumber\\
&&
\qquad-
2\big\ipbra
\big(\ncp_{{\bf k}}+({T}/{T_p})\,{\smear}_{{\bf k}}\,\nop_{{\bf k}}^{0}\big)
\nop^{\phantom{0}}_{-{\bf k}}
\big\ipket
\Big).
\end{eqnarray}
Here, $\int_{\bf k}$ is shorthand for $\int d^{D}k/(2\pi)^{D}$,
$k^{2}$ is the squared length of the vector ${\bf k}$,
and the $R_{{\bf k}}$ is the Fourier transform
$\int d^{D}r\,R({\bf r})\,\exp(i{\bf k}\cdot{\bf r})$.
In addition, curly brackets---as in $\{{\mathbf S}\,{\mathbf S}^{\prime}\}$---indicate
the trace of the product of the tensors ${\mathbf S}$ and ${\mathbf S}^{\prime}$, i.e.,
$\sum_{d,d^{\prime}=1}^{D}S_{dd^{\prime}}S^{\prime}_{d^{\prime}d}$.
Furthermore, $\LDLa$ characterizes the aligning tendencies of nematic freedoms; its value can be obtained by a complementary microscopic calculation as $J^2/T$ (see Ref.~\cite{ref:microscopic}), and $\stiff$ is the Frank constant for nematic order.
Moreover, $t$ is the reduced measurement temperature,
defined to be $(T-T^{\ast})/T^{\ast}$,
where $T$ is the measurement temperature (i.e., the temperature at which the system is maintained, in equilibrium, long after the cross-linking process), and $T^{\ast}$ is the spinodal temperature for the spatially homogeneous isotropic-to-nematic transition.
We also introduce the temperature $T_p$ of the equilibrium state into which cross-links are instantaneously created, where $p$ stands for preparation.
The occurrence of two temperatures, $T$ and $T_p$, stems from the fact that elastomers and related systems are characterized by not one but two statistical ensembles.  One, which we call the {\it preparation ensemble\/}, provides a statistical description of the random (non-equilibrating, unmeasured) freedoms $\nop^{0}$ that characterize the local alignment  immediately prior to cross-linking.  The other describes the equilibrium state of the system long after cross-linking was done, via the statistics of the equilibrating variables $\nop$; we call it the {\it measurement ensemble\/}.

The free energy~(\ref{eq:landau}) consists of two terms.
The first two elements of the first term constitute the familiar Landau-de~Gennes free energy; they describe the free-energetic cost of inducing nematic alignment from the unaligned state.
The second term incorporates what we have described above, viz.,
the influences of
(i)~the configuration of the rod-like constituents at the instant of cross-linking, via $\nop^{0}$,
together with
(ii)~the memory-independent random field $\ncp$ caused by the localized polymers post cross-linking.
The contribution to $F/T$ involving $\nop^0$ carries a factor $(J/T_p)(G\ell^{2}/\loclen^2)^{2}(J/T)$.
The two temperature factors show that the network is better able to store a given pattern $\nop^0$ when the preparation temperature $T_p$ is lower and, similarly, better able to elicit $\nop^0$ from $\nop$ the lower the measurement temperature $T$.
Taken together, the two terms in $F$ are minimized by the most probable nematic configuration $\widetilde{\nop}$, which is

\begin{eqnarray}
\widetilde{\nop}_{\bf k}=
\frac{\ncp_{\bf k}+
({T}/{T_p})\,
{\smear}_{\bf k}\,
\nop^{0}_{\bf k}}
{\LDLa t+\stiff k^{2}+\smear_{\bf k}}.
\end{eqnarray}
The third element in the first term of the free energy~(\ref{eq:landau}) is a new and central element.  It encodes the essential physical difference between nematic fluids and nematic elastomers, viz., the elastomer's possession of a network that is localized randomly and fluctuating thermally, and is, furthermore, liquid-like at sub-localization-length scales and solid-like at larger scales.  As a result of this solidness at larger scales, creating a pattern of nematic alignment (either by making relative translations of localized nematogens whilst maintaining their orientations, or by locally reorienting the nematogens) carries an additional free-energy cost, compared to nematic fluids, owing to the need to compete with localization forces.
In a similar vein,
by completing the square with respect to the first and second terms in Eq.~(\ref{eq:landau}), and thus arriving at the form
\begin{equation}
\label{eq:nematicdepartures}
\frac{1}{2}
\int_{{\bf k}}
\big(\LDLa t+\stiff k^{2}+{\smear}_{\bf k}\big)
\big\ipbra
\big(\nop_{ {\bf k}}-\widetilde{\nop}_{ {\bf k}}\big)
\big(\nop_{-{\bf k}}-\widetilde{\nop}_{-{\bf k}}\big)
\big\ipket
\end{equation}
(up to a non-thermally fluctuating term),
the third element in Eq.~(\ref{eq:nematicdepartures}) is seen to give a nonlocal free-energy cost of creating a departure from the nematic pattern $\widetilde{\nop}_{\bf k}$.  This cost arises because the network mediates additional nematic-nematic interactions.  Moreover, these nonlocal interactions are not independent of the memory-independent random field $\ncp$ that acts to produce local nematic alignment, as this field also originates with the network.
Thus, it is relatively costly, energetically, to create a nematic departure from $\widetilde{\nop}$ that is essentially uniform over a lengthscale rather larger than $\loclen$, at which scale the solidness of the network becomes pronounced.  On the other hand, the cost is relatively mild if the departure varies only over some lengthscale rather shorter than $\loclen$, where the system has a more liquid-like character.

Various averaged diagnostics of the system involving the local nematic order $\nop({\bf r})$ may be considered via $F$.  These averages come in two types.  First, there are disorder-averaged quantities (denoted by $[\cdots]$), by which we mean quantities averaged over suitably distributed $\ncp$ and $\nop^{0}$.  Second, there are thermally averaged quantities (denoted by $\langle\cdots\rangle$), by which we mean quantities averaged over the measurement ensemble, i.e., the familiar statistical-mechanical ensemble of the equilibrating freedoms, in this case $\nop$.  Here, we focus on two particular diagnostics of nematic elastomers.  The first, the thermal fluctuation correlator $\mycorrel^{T}$, defined via
\begin{subequations}
\label{eq:coredefs}
\begin{equation}
\label{eq:thermcorrdef}
\mycorrel^{T}({\bm r}-{\bm r}^{\prime})
\equiv
\Big[
\Big\langle
\Big\ipbra
\big(\nop({\bm r})-\langle\nop({\bm r})\rangle\big)
\big(\nop({\bm r}^{\prime})
-\langle\nop({\bm r}^{\prime})\rangle\big)
\Big\ipket
\Big\rangle
\Big],
\end{equation}
characterizes the strength of the thermal fluctuations of the nematic alignment away from the local mean value as well as the spatial range over which these fluctuations are correlated.  Inter alia, through its range, $\mycorrel^{T}$ can be used to detect the occurrence of a continuous phase transition.
The second is the glassy  fluctuation correlator $\mycorrel^{G}$, defined via
\begin{equation}
\label{eq:glasscorrdef}
\mycorrel^{G}({\bm r}-{\bm r}^{\prime})
\equiv
\Big[
\Big\ipbra
\big\langle\nop({\bm r}         )\big\rangle
\big\langle\nop({\bm r}^{\prime})\big\rangle
\Big\ipket
\Big].
\end{equation}
\end{subequations}
This is a diagnostic of any randomly frozen (i.e., time-persistent) nematic order present.  In particular, its value at the origin, $\mycorrel^{G}(\rv)|_{\rv=\bm{0}}\,$, is the nematic analog of the Edwards-Anderson order parameter for spin glasses~\cite{Edwards-Anderson-SpinGlass-1975}, and measures the magnitude of the local frozen nematic order; hence the name \textit{glassy correlator}.  Moreover, how $\mycorrel^{G}(\rv)$ varies with the separation ${\bm r}$ determines the spatial extent of regions that share a roughly common nematic alignment.  For isotropic liquids the glassy correlator automatically vanishes, but for IGNEs it is nonzero.
The correlators  $\mycorrel^{T}$ and $\mycorrel^{G}$, which diagnose the measurement ensemble, give valuable information about the physical structure of nematic elastomers~\cite{ref:cartesiancor}.  (Other correlators, including ones that, in addition, probe the preparation ensemble, can also be considered.  A broad account of the issues that result from the presence of a multiplicity of ensembles and the correlators that diagnose them will be given in a forthcoming paper~\cite{ref:DErevisit}.)

One could also consider the disorder-averaged quantity $[\langle \nop(\rv) \rangle]$; it, however, vanishes, owing to the macroscopic isotropy of the post-cross-linking state.
On the other hand, the thermal average of the local order parameter for a specific realization of the quenched disorder $\langle \nop(\rv)\rangle$ is maintained at a nonzero, time-persistent, random value, which we shall compute shortly.
This nonzero value is the result of the partial trapping, by the network, of the orientational randomness $\nop^{0}$ present at the instant of cross-linking, together with the memory-independent random field $\ncp$ of the network, post cross-linking.
The free energy~(\ref{eq:landau}) is quadratic in $\nop$, and therefore the computation of $\langle \nop\rangle$ and $\mycorrel^{T}$ using the statistical weight $\exp(-F/T)$ is elementary, yielding
\begin{eqnarray}
\!\!\!\!\!\!\!\!\!&&\big\langle \nop_{-{\bf k}}\big\rangle
=
\frac{\ncp_{\bf k}\!+\!(T/T_p){\smear}_{\bf k}\nop^{0}_{\bf k}}
{\LDLa t\!+\!\stiff k^{2}\!+\!{\smear}_{\bf k}},
\\
\!\!\!\!\!\!\!\!\!&&\Big\langle\!
\Big\{\!
\big(\nop_{{\bf k}} - \langle \nop_{{\bf k}}\rangle \big)
\big( \nop_{{\bf k}^{\prime}} - \langle \nop_{{\bf k}^{\prime}}\rangle \big)
\!\Big\}
\!\Big\rangle
=
\frac{T\mu_D\,\delta_{{\bf k}+{\bf k}^{\prime},{\bf 0}}}
{\LDLa t + \stiff k^{2} + {\smear}_{\bf k}}.
\end{eqnarray}
Here, $\mu_D\equiv (D-1)(D+2)/2$ counts the number of degrees of freedom of $\nop$ and takes the value $5$ for $D=3$.
Note that we have chosen units in which Boltzmann's constant has the value unity.

To perform the average over the quenched random variables $\ncp$ and $\nop^{0}$ we must adopt a model for their statistics that is consistent with the physical origin each has.  The choice we make is that $\ncp$ and $\nop^{0}$ are independent, Gaussian-distributed random fields, with vanishing means and non-vanishing variances, the latter being given by
\begin{subequations}
\label{eq:disorderstats}
\begin{eqnarray}
\big[\big\{\nop^{0}_{{\bf k}}\,\nop^{0}_{{\bf k}^{\prime}}\big\}\big]
&=&
T_p\frac{\mu_D\,\delta_{{\bf k}+{\bf k}^{\prime},{\bf 0}}}{\LDLa^{0}t_{p}+\stiff^0 k^{2}},
\\
\big[\big\{\ncp_{{\bf k}}\,\ncp_{{\bf k}^{\prime}}\big\}\big]
&=&
T\,{\smear}_{\bf k}\,
\delta_{{\bf k}+{\bf k}^{\prime},{\bf 0}}\,.
\end{eqnarray}
\end{subequations}
Here, $\LDLa^{0}$ is the preparation-ensemble counterpart to $\LDLa$, and a complementary microscopic calculation shows that its value is given by $J^2/T_p$ (see Ref.~\cite{ref:microscopic}).  Similarly, $\stiff^0$ is the preparation-ensemble counterpart to $\stiff$.
The statistics of $\nop^{0}$ depends on $t_{p}$
[i.e., the parameter that encodes the temperature $T_{p}$ of the preparation ensemble via $t_{p}\equiv(T_{p}-T^{\ast})/T^{\ast}$];
it does not depend on $\smear$, because $\smear$ encodes the physics of random
but imperfect spatial localization, and this only comes into being post cross-linking.
(The {\it impact\/} of $\nop^{0}$ {\it does\/} depend on $\smear$, as $\smear$ controls
the relaxation of $\nop$ from $\nop^{0}$ to its equilibrium value, post cross-linking.)\thinspace\
By contrast, the statistics of $\ncp$ {\it does\/} depend on $\smear$;
this is because $\smear$ characterizes the typical value of the memory-independent random field that
results from the random imperfect spatial localization of the polymers constituting the network.
In view of their distinct origins it is natural that $\ncp$ and $\nop^{0}$ are statistically uncorrelated.
However, it is also natural, at least heuristically, that the $\smear$ that characterizes the \lq\lq orientational caging\rq\rq\ induced by the network (via $\ncp$) is the same $\smear$ that determines the fidelity with which the network preserves the orientational order present immediately post cross-linking (i.e., $\nop^{0}$).  It is natural because localization that is sharper and more widespread (i.e., involves a larger localized fraction) both {\it creates\/} more intense network-induced orientational caging and {\it enhances\/} the trapping-in of the local nematic order present immediately post cross-linking.  Our physical expectation, borne out by a complementary  microscopic analysis (see Ref.~\cite{ref:microscopic}), is that such strengthening of the localization would more strongly enhance memorization than orientational caging.  This expectation is consistent with the phenomenological choice presented here, in which the corresponding contributions to the random anisotropy field, Eq.~(\ref{eq:anisotropy}), scale as $\sqrt{\smear}$ for the caging (i.e., $\ncp$) part and ${\smear}$ for the \lq\lq memorization\rq\rq\ (i.e., $\nop^0$) part.

Returning to the disorder-averaged diagnostics---the mean value $[\langle \nop\rangle]$ and the correlators $\mycorrel^{T}$ and $\mycorrel^{G}$---we complete their computation by using the statistics of the quenched disorder, Eqs.~(\ref{eq:disorderstats}), to arrive at
\begin{subequations}
\label{eq:correl}
\begin{eqnarray}
&&\!\!\!\!\!\!\!\!\!\!\!\!\!\!\!
\big[\big\langle \nop_{{\bf k}}\big\rangle\big]=
0,
\\
&&\!\!\!\!\!\!\!\!\!\!\!\!\!\!\!
\mycorrel^{T}_{\bf k}=
T\mu_D\,\frac{1}{\LDLa t + \stiff k^{2}+{\smear}_{\bf k}},
\label{eq:correlT}
\\
&&\!\!\!\!\!\!\!\!\!\!\!\!\!\!\!
\mycorrel^{G}_{\bf k}=
T\mu_D\,\frac{
\frac{T}{T_p} (\LDLa^{0}t_p+\stiff^0 k^{2})^{-1}\vert\smear_{\bf k}\vert^{2}
+\smear_{\bf k} 
}
{\left(\LDLa t\!+\!\stiff k^{2}\!+\!{\smear}_{\bf k}\right)^{2}}.
\label{eq:correlG}
\end{eqnarray}
\end{subequations}
Having computed the correlators $\mycorrel^{T}$ and $\mycorrel^{G}$, we now set about using them to study how the presence of a network modifies the organizational behavior of nematic freedoms.  To do this, we first note that there are two emergent lengthscales present in IGNEs: (i)~the typical localization length, $\xi_L$, quantifying the sharpness of localization of polymers belonging to the network, and (ii)~the {\it intrinsic nematic correlation length\/}, $\xi_N$ [$\equiv \sqrt{\stiff/\LDLa t}$], describing the range over which nematic freedoms would be correlated if there were no network present.  On the other hand, we have the strength of the memory-independent random field $\ncp$, which is characterized by $\sqrt{\smear_{\bf 0}}$.  In what follows, we shall study the effects of these three parameters on $\mycorrel^{T}$ and $\mycorrel^{G}$ at two preparation temperatures, one {\it far\/} above $T^*$ and the other {\it just\/} above it.


\renewcommand{\arraystretch}{1.5}
\begin{center}
\begin{table}[h]
\begin{ruledtabular}
\begin{tabular}{c|l|r}
Disorder strength&Weak ($\smear_{\bf 0}<\smear_{c}$)&Strong ($\smear_{\bf 0}>\smear_{c}$)\\
\hline
$\xi_{T,o}^{2}$&$\infty$&
$\frac{1}{2}\xi_{L}^{2}/\ln(\smear_{\bf 0}/\smear_{c})$
\cr
$\xi_{T,d}^{2}$&
$\xi_{N}^{2}\frac{1-({\smear_{\bf 0}}/{\smear_{c}})}{1+(\smear_{\bf 0}/{\LDLa t})}$&
$\sim \xi_{L}^{2} / \big( 1 + \frac{\xi_L^2}{2 \xi_N^2} \big)$
\cr
$\xi_{G,o}^{2}$&
$\infty$&
$\sim\frac{1}{2}\xi_{L}^{2}/\ln(\smear_{\bf 0}/\smear_{c})$
\cr
$\xi_{G,d}^{2}$&
$\frac{1}{2}\xi_{L}^{2}+
2\xi_{N}^{2}\frac{1-({\smear_{\bf 0}}/{\smear_{c}})}{1+(\smear_{\bf 0}/{\LDLa t})}$&
$\sim \frac{1}{2} \xi_L^2$
\cr
$\mycorrel^{G}({\bf r}={\bf 0})$&
$\sim T\smear_{\bf 0}\sqrt{\frac{\pi}{8\LDLa t \stiff^3}}\frac{\xi_N}{\xi_L}$&$\sim T\frac{\pi}{4}\sqrt{\smear_{\bf 0}/\stiff^3}$
\end{tabular}
\end{ruledtabular}
\caption{Values of the correlation lengthscales ($\xi_{T,d}$ and $\xi_{G,d}$), the oscillation wavelengths ($\xi_{T,o}$ and $\xi_{G,o}$), and the intensity of local nematic alignments [$\mycorrel^{G}({\bf r}={\bf 0})$] in the weak-- and strong--disorder regimes for the case of IGNEs crosslinked at temperatures far above $T^*$.}
\label{table:summary}
\end{table}
\end{center}

First we consider the behaviors of $\mycorrel^{T}$ and $\mycorrel^{G}$ for systems prepared at some temperature far above $T^*$, so that any local nematic order present immediately post cross-linking (and thus available for trapping in) is spatially correlated only over distances far shorter than the typical localization length $\xi_L$; see Table~\ref{table:summary}.  This implies that the local nematic order arising from $\nop^0$ would be heavily \lq\lq washed out\rq\rq\ by thermal fluctuations of the network.  Thus, in this regime, the dominant contribution to the trapped-in local nematic order originates in the memory-independent random field, $\ncp$, and concomitantly, the contribution arising from $\nop^{0}$ may be neglected.

$\mycorrel^T$ and $\mycorrel^G$ exhibit qualitatively distinct behaviors in two regimes, depending on the strength of the random field.  For $\smear_{\bf 0} < \smear_c$ (where $\smear_c\equiv 2\stiff/\xi_{L}^{2}$---the {\it weak disorder regime\/}), $\mycorrel^{T}$ and $\mycorrel^{G}$ {\it simply decay\/} with increasing real-space separation.  More specifically, by examining their small wave-vector behaviors we ascertain that the respective associated correlation lengths $\correllength_{T,d}$ and $\correllength_{G,d}$ have the values given in Table~\ref{table:summary}.  We see from the behavior of $\xi_{T,d}$ the physically reasonable result that the random network, with its thermal fluctuations, serves to shorten the nematic thermal fluctuation correlation length from the value it would have absent the network, a phenomenon that a conventional (i.e., non-thermally fluctuating) random-field approach would not capture.
As for $\xi_{G,d}^{2}$, it comprises two parts.
One ($\propto\xi_{T,d}^{2}$) arises from the nematic thermal correlations;
the other ($\propto\xi_{L}^{2}$) comes from the local aligning effect exerted by the cage.
The fact that $\xi_{G,d}$ increases with $\xi_{L}$ does not mean that a more weakly cross-linked network (for which $\xi_{L}$ would be larger) aligns the nematogens more effectively.  Whilst the {\it lengthscale\/} of aligned regions $\xi_{G,d}$ may increase, the {\it magnitude\/} of $\mycorrel^{G}$, which governs the intensity of the alignment locally, decreases, as can be seen from Table~\ref{table:summary}.

On the other hand, the simple decay of the correlators at weak disorder can give way to oscillatory decay at strong disorder, i.e., $\smear_{\bf 0}/\smear_{c} > 1$.
$\mycorrel^{T}$ has such behavior regardless of $T$, whereas
$\mycorrel^{G}$ has it only for sufficiently small $T$.
The oscillation wavelengths $\xi_{T/G,o}$ are determined via the radii of the shells in wave-vector space
on which the corresponding correlators are maximal.
Thus, we arrive at an explicit (and, notably, $T$-independent) formula
$\xi_{T,o}=\xi_L/\sqrt{2\ln( \smear_{\bf 0}/\smear_c )}$ and
an implicit one for $\xi_{G,o}$, viz.,
\begin{equation}
\label{eq:nonlinear}
1 + (\xi_N/\xi_{G,o})^2 + 4(\xi_N/\xi_L)^2 - (\smear_{\bf 0}/\LDLa t) e^{-\xi_L^2/2\xi_{G,o}^2}=0.
\end{equation}
The value of $\xi_{T/G,d}$ in this strong-disorder regime, given in Table~\ref{table:summary}, is estimated via the widths of the peaks of $\mycorrel^{T/G}$. Upon decreasing $\xi_{L}$ at fixed $\xi_N$, the value of $\xi_{T,d}$ tends to $\xi_{L}$ from above, indicating that the network is limiting the range over which the thermal nematic fluctuations are correlated; on the other hand, $\xi_{G,d}$ remains at the scale of $\xi_L$, indicating that the range of coherent nematic alignment is circumscribed by the network's typical localization length.  Concomitantly, there is a growth in the intensity of the local alignment $\mycorrel^{G}({\bf r}={\bf 0})$.

Having considered the behaviors of $\mycorrel^{T}$ and $\mycorrel^{G}$ for systems prepared at high temperatures, we now consider the corresponding behavior for systems prepared at a temperature just above $T^*$, so that the local nematic order present immediately post cross-linking is spatially correlated over distances larger than the typical localization length of the elastomeric network.
As one can see from Eq.~(\ref{eq:correlT}), the behavior of $\mycorrel^{T}$ is unchanged, undergoing simple decay in real space at weak disorder and oscillatory decay at strong disorder.
Conversely, $\mycorrel^{G}$ exhibits behavior qualitatively different from that of a system prepared at high $T_p$, because it now receives its dominant contribution from the memorization of $\nop^0$.  Specializing to $t\approx t_p$, we see from Eq.~(\ref{eq:correlG}) that $\mycorrel^{G}$ is approximately given by
\be
\mycorrel_{\bf k}^G \approx \mu_D\left( \frac{T}{T_p} \right)^2 \frac{T_p}{\LDLa^0 t_p + \stiff^0 k^2},
\ee
which is proportional to the correlator of the thermal nematic fluctuations immediately post cross-linking. This indicates that the pattern of these thermal fluctuations has been faithfully memorized by the network.

To conclude, we have proposed a physically motivated phenomenological model to describe isotropic-genesis nematic elastomers. Our approach enables us to characterize the random nematic alignment present, post cross-linking, in terms of both the partial memorization by the network of the nematic alignment present at the genesis of the network and the structural anisotropy presented by the network itself.  Our approach also reveals that networks that have sufficiently strong aligning capabilities exhibit a novel, oscillatory-decaying type of spatial correlation of nematic alignments. As we shall explore in detail in two companion papers~\cite{ye2,ref:microscopic}, the physical origin of these oscillatory correlations is the thermally compliant nature of the random fields in these systems.  The nematogens subject to similar random anisotropic environments tend to aggregate whilst those subject to different anisotropic environments tend to segregate.
In addition to its specific results, the present work underscores the necessity of broadening the concept of a quenched random field to incorporate not only the traditional, \lq\lq frozen\rq\rq\ type, which does not fluctuate thermally~\cite{terentjev1,terentjev2,ImryMa,radzihovsky,feldman}, but also the type in which the frozen character of the random field is present only at longer lengthscales, fading out as the lengthscale progresses to below a characteristic localization length, owing to the thermal position fluctuations of the network constituents.

\begin{acknowledgments}
We thank
Kenji Urayama,
Tom Lubensky,
Leo Radzihovsky
and Mark Warner
for informative discussions.
This work was supported by the National Science Foundation via grant no.~DMR09-06780,
the Institute for Complex Adaptive Matter, and Shanghai Jiao Tong University.
\end{acknowledgments}

\newpage
\setcounter{secnumdepth}{-1}


\begin{thebibliography}{99}
\bibitem{ref:cartesiancor}
Although we are focusing on the Cartesian scalar aspects of the correlators, it is straightforward to reconstruct the full structure of the corresponding fourth-rank Cartesian tensors, by appending suitable isotropic tensor factors constructed from Kronecker deltas.
\end{thebibliography}
\end{document}